\documentstyle[12pt]{article}
\normalsize
\def\sp{~~~~~}

\def\a{\alpha}
\def\b{\beta}

\def\d{\delta}
\def\e{\epsilon}
\def\f{\phi}
\def\g{\gamma}
\def\h{\eta}

\def\m{\mu}
\def\n{\nu}

\def\p{\pi}

\def\r{\rho}
\def\s{\sigma}

\def\x{\xi}

\def\D{\Delta}

\def\G{\Gamma}

\def\cd{{\cal D}}

\def\cl{{\cal L}}

\def\co{{\cal O}}

\def\Rt{\rightarrow}

\def\pa{\partial}

\def\iff{\leftrightarrow}

\def\bar#1{\overline{#1}}

\def\Hat#1{\rlap{\kern.10em$\widehat{\phantom G}$}#1}
\def\HAt#1{\rlap{\kern.05em$\widehat{\phantom G}$}#1}

\def\cap#1{\rlap{\kern.1em$\widehat{\phantom{G\vrule height.8em}}$}#1{}}
\def\Cap#1{\rlap{\kern.05em$\widehat{\phantom{G\vrule height.8em}}$}#1{}}

\let\oldtheequation=\theequation
\def\doteqs#1{\setcounter{equation}{0}
            \def\theequation{{#1}.\oldtheequation}}
\newcounter{sxn}
\def\sx#1{\addtocounter{sxn}{1} \bigskip\medskip \goodbreak
\noindent{\large\bf\centerline{\thesxn.~~#1}} \nobreak \medskip}
\def\sxn#1{\sx{#1} \doteqs{\thesxn}}

\newcounter{axn}

\def\br{\begin{thebibliography}{abc}}
\def\er{\end{thebibliography}}
\def\rf{\bibitem}
\date{}

\begin{document}
\bibliographystyle{unsrt}
\footskip 1.0cm
\thispagestyle{empty}
\setcounter{page}{0}
\begin{flushright}
SU-4240-519\\
September 1992\\
\end{flushright}
\vspace{10mm}
\centerline {\bf\LARGE ROLE OF LIGHT VECTOR MESONS IN THE}
\vspace{5mm}
\centerline {\bf\LARGE HEAVY PARTICLE CHIRAL LAGRANGIAN}
\vspace*{5mm}
\centerline {\large J. Schechter  \small and \large A. Subbaraman}
\vspace*{5mm}
\centerline {\it Department of Physics, Syracuse University,}
\centerline {\it Syracuse, NY 13244-1130}
\vspace*{25mm}
\normalsize
\centerline {\large Abstract}
\vspace*{5mm}
We give the general framework for adding ``light'' vector particles to the
heavy hadron effective chiral Lagrangian.  This has strong motivations both
from the phenomenological and aesthetic standpoints.  An application to the
already observed $D\rightarrow \bar{K}$* weak transition amplitude is
discussed.

\newpage

\baselineskip=24pt
\setcounter{page}{1}
\sxn{Introduction}
Recently there has been a lot of interest in the heavy quark (or Isgur-Wise)
symmetry \cite{Wis} which pertains to a rigorous limit of QCD in which old
fashioned quark model results may be applied.  This limit corresponds to
keeping the four-velocity, $V_\mu$ of the heavy quark fixed while taking its
mass, $M$ to infinity.  A natural application of this approach is to the chiral
interactions of the heavy particles with ``soft'' pions and kaons.  Indeed a
number of interesting papers \cite{Wi2,Yan,Bar,Cho,Jen}
have already appeared.  The resulting effective Lagrangians can
be used to
relate amplitudes for processes with a fixed number and type of heavy quarks
but with any number of soft pseudoscalars.  For example the amplitude for
$D^{+}\rightarrow e^{+} \nu_e$ is related to the amplitude for $D^0
\rightarrow
K^-e^+\nu_e$ in the soft $K^-$ region.  Continuing, these amplitudes are
related to that for $D^+ \rightarrow K^-\pi^0e^+\nu_e$ with soft $K^-$ and
$\pi^0$.  It would be very interesting to compare such a relation with
experiment.  Unfortunately, on consulting the Review of Particle Properties we
learn \cite{Par} that ``it is generally agreed that the $\bar{K}\pi
e^+\nu_e$ decays of the $D^{+}$ and $D^0$ are
dominantly $\bar{K}^{*}e^+\nu_e$''.
This is not very surprising since it is known from low energy physics that two
pseudoscalars often prefer to make their appearance as a vector meson.  In the
future it will undoubtedly be possible to disentangle the non-resonant two
pseudoscalar piece.  But this example provides a strong motivation for
including the light vector mesons in the formulation of the heavy particle
effective Lagrangian.  We will begin the investigation of the heavy particle
effective chiral Lagrangian with vectors in the present paper.  The application
to $D^{+} \rightarrow \bar{K}^{*0}e^{+}\nu_e$ will also be
discussed.

In section 2 we will discuss the derivation of the non-interacting part of the
heavy meson Lagrangian in order to set down our notation and make some points
which will be useful later on.  Section 3 contains a brief treatment of the
chiral Lagrangian of light pseudoscalars and vectors as well as the
interactions of these fields with the heavy mesons.  Compared to the heavy
meson chiral Lagrangian with only light pseudoscalars there is now a modified
chiral covariant derivative as well as a characteristic new interaction term.
We will employ a phase convention for the ``heavy meson fields'' which is
convenient for making contact with ``ordinary'' meson fields and verifying the
CP invariance of the theory.  In section 4 we will give the leading chiral
covariant expression for the weak current and, apply it, in section 5, to the
soft light meson regions of the $D^0 \rightarrow K^-$ and $D^{+}
\rightarrow
\bar{K}^{*0}$ transition matrix elements.
\sxn{Derivation of non-interacting Lagrangian}
Let us denote the heavy mesons associated with each heavy flavor as being made
out of the heavy quark (rather than anti quark); symbolically
\begin{equation}
{\rm heavy~meson~field} \sim \bar{q}_{\rm light}q_{\rm heavy}.
\end{equation}
Since there are three light flavors, (2.1) should be regarded as a three
component {\it row} vector for each heavy flavor.  In the presently known cases
we thus have the experimental pseudoscalar objects
$(D^0,D^{+}, D^{+}_s)$ and
$(\bar{B}^-,\bar{B}^0,\bar{B}^0_s)$.

For our purpose it will be instructive to derive a heavy meson field effective
Lagrangian directly from an ordinary field effective Lagrangian, in analogy ot
the treatment \cite{Geo} \cite{Wis} of the heavy quark effective
Lagrangian.  Let us first consider the non-interacting terms
\begin{equation}
\cl_{\rm free}(P)=-\partial_{\mu}P\partial_\mu\bar{P}-M^2P\bar{P},
\end{equation}
for the heavy pseudoscalar (row vector) field $P(x)$ of mass $M$.  Note that
$\bar{P} \equiv P^\dagger$ and that we are employing the ``Euclidean'' metric
convention with $x_4=it$.  In order to implement the basic idea that deviations
from straight line motion with 4-velocity $V_\mu$ of the heavy meson be small
we make the change of variables
$$
P=e^{iMV\cdot x}P'
$$
\begin{equation}
\bar{P}=e^{-iMV\cdot x}\bar{P}'.
\end{equation}
$V_\m$ should be considered as fixed.  Furthermore, in the free field expansion
\begin{equation}
P=\sum_{\underline K}\frac{1}{\sqrt{2E_{\underline K}V}}\left(a_{\underline
K}e^{iK\cdot x}+b_{\underline K}^{\dag}e^{-iK\cdot x}\right),
\end{equation}
we are considering that the anti-particle operators $b_{\underline K}$ should
be neglected.  (In the interacting theory, the anti-particles won't be excited
as $M \rightarrow \infty$).  Substituting (2.3) into (2.2) yields
\begin{equation}
\cl_{\rm free}(P')=-iMV_\mu
P'\buildrel\leftrightarrow\over{\partial_\mu}\bar{P}'-\partial_\mu
P'\partial_\mu\bar{P}'.
\end{equation}
Notice that the terms of order $M^2$ have cancelled out.  The second term in
(2.5) is negligible as $M\rightarrow\infty$ so we simply have in the heavy
quark limit
\begin{equation}
\cl_{\rm free}(P')=-2iMV_\mu P'\partial_\mu\bar{P}'.
\end{equation}
This can be simplified further by redefining
\begin{equation}
P''=M^{1/2}P'
\end{equation}
to give a form in which the mass independence is manifest,
\begin{equation}
\cl_{\rm free}(P'')=-2iV_\mu P''\partial_\mu \bar{P}''
\end{equation}
However, $P''$ has the non-canonical dimension $\frac{3}{2}$.

Let us next consider the heavy vector field $Q_\mu$, which is relevant because
it belongs \cite{Wis} to the same heavy spin multiplet as $P$.  The free
Lagrangian in terms of ordinary spin one fields is
$$
\cl_{\rm free}(Q)=-\frac{1}{2}(\pa_{\m}Q_{\n}-\pa_{\n}Q_{\m})
(\pa_{\m}\bar{Q}_{\n}-\pa_{\n}\bar{Q}_{\m})-M^2Q_{\m}\bar{Q}_{\m},
$$
\begin{equation}
\bar{Q}_{\mu}=(-1)^{\d\m4} Q^{\dag}_{\m}.
\end{equation}
The transformation
$$
Q_{\m}=e^{iMV\cdot x} Q_{\m}'
$$
\begin{equation}
\bar{Q}_{\m}=e^{-iMV\cdot x}\bar{Q}_{\m}'
\end{equation}
then yields the ``small oscillation'' Lagrangian,
\begin{equation}
\cl_{\rm free}(Q')=-2iMV_{\n}Q_{\m}'\pa_{\n}\bar{Q}_{\m}',
\end{equation}
in which the subsidiary condition $V_{\m}Q_{\m}'=0$ was imposed and a term
negligible as $M\Rt\infty$ was dropped.

Note that (2.6) and (2.11) both have the same structure.  This is to be
expected by the heavy quark symmetry and can be made \cite{Bjo} \cite{Wis}
manifest by amalgamating $P'$ and $Q_{\m}'$ into a single ``heavy quark''
field,
$H$:
$$
H=\left(\frac{1-i\g\cdot V}{2}\right)(\h\g_5 P'+i\g\cdot Q'),
$$
\begin{equation}
\bar{H}\equiv \g_4 H^{\dag} \g_4=(-\h^{*}\g_5 \bar{P}'+i\g\cdot \bar{Q}')
\left(\frac{1-i\g\cdot V}{2}\right).
\end{equation}
Here $H$ is a $4\times 4$ matrix in the Dirac spinor space and the coefficients
of $P'$ and $Q_{\m}'$ are the kinematical operators which respectively project
out
the pseudoscalar and the vector combinations from $\bar{q}_{\rm light}
q_{\rm heavy}$.  $\h$ is an arbitrary phase which we will choose as
\begin{equation}
\h=i,
\end{equation}
for a reason to be discussed later.  In contrast, $\h$ is chosen to be purely
real in ref. 2.  Using (2.12), the sum of (2.6) and (2.11) can be compactly
written as:
\begin{equation}
\cl_{\rm free}(P',Q')=iMV_\m Tr(H\pa_{\m}\bar{H}),
\end{equation}
where the trace refers to the $4\times 4$ Dirac space.  There is also an
implied summation in the light flavor space since $H$ is a row vector and
$\bar{H}$ is a column vector.  The use of the $H$ field evidently \cite{Wis}
guarantees the invariance under heavy quark spin transformations (in the Dirac
space): $H\Rt SH, \bar{H}\Rt\bar{H}S^{-1}$.

Since (2.14) represents the heavy quark limit of (2.2) plus (2.9) one might
think that the sum of (2.2) plus (2.9) before taking the limit should be more
compactly written using an $H$ defined in terms of $P$ and $Q_{\m}$ (rather
than
$P'$ and $Q_{\m}'$) as
$$
\frac{1}{2}Tr(\pa_{\m}H\pa_{\m}\bar{H})+\frac{1}{2}M^2Tr(H\bar{H}).
$$
This expression is not however consistent even though it does reproduce (2.14)
in the heavy quark limit after the substitutions (2.3) and (2.10) are made.
The
reason is that it gives a vector kinetic term $-\pa_{\m}Q_{\n}\pa_{\m}
\bar{Q}_\n$
which (unlike (2.9)) is well-known to lead to a Hamiltonian unbounded from
below.  This example illustrates the danger of using $H$ outside the heavy
quark regime.

We remark that even though (2.14) [as well as the interacting analogs to be
discussed later] is very compact it is not any more difficult to use the sum of
(2.2) and (2.9) for practical calculations.  This is because the Feynman rules
for ordinary mesons are very well known and we can just substitute for the
heavy quark momentum $K_{\m}$, $K_{\m} =MV_{\m}+K_{\m}'$ at the end of the
calculation.
{}From this point of view the heavy quark symmetry just tells us to put the $P$
and $Q_\m$ masses equal and to equate certain coefficients of the interaction
Lagrangian.  Certainly it is useful to keep both approaches in mind.  In one
respect the use of the ordinary fields might actually appear more convenient.
That is the case when we want to consider the heavy limit for mesons containing
heavy anti-quarks.  The ordinary fields contain both quark and antiquark
operators as in (2.4) so the calculation can be done using standard techniques.
However, by construction, the heavy quark Lagrangian makes no reference to
antiquarks.  Of course this is not a big problem and one can define a heavy
anti-particle Lagrangian in a similar way.  We would then like to choose the
heavy anti-particle fields to be related to the heavy particle fields in the
same way that the ordinary fields describe both particles and anti-particles.
What this amounts to is choosing a phase convention so that the right hand
sides of (2.3) and (2.10) transform in the same way under charge conjugation as
the left hand sides.  That turns out to be the reason for our choice (2.13).
\sxn{Chiral Interaction Terms}
First consider the three light (current) quarks, $q$.  Under a chiral
transformation
\begin{equation}
q_L\rightarrow U_Lq_L,\sp q_R\rightarrow U_Rq_R,
\end{equation}
where $U_L$ and $U_R$ are $3\times 3$ unitary matrices.  The chiral matrix
$U=\exp[2i\f/F_{\p}]$, where $\f$ is the $3\times 3$ matrix of pseudoscalars
and
$F_\p\simeq 132~{\rm MeV}$, is constructed \cite{Cro} in such a way that the
interaction term $\bar{q}_L Uq_R+{\rm h.c.}$ is invariant.  This implies
$U\Rt U_L UU^{\dag}_R$.  The interaction term is converted to a light
quark ``constituent'' mass term by the change of variables $q_L=\x\tilde{q}_L$,
$q_R=\x^{\dag}\tilde{q}_R$ with $\x \equiv U^{1/2}$.  The transformation
property of
$U$ implies \cite{Cal}
\begin{equation}
\x\Rt U_L\x K^{\dag}=K\x U^{\dag}_R,
\end{equation}
where the unitary matrix $K$ depends on $U_L$, $U_R$ as well as $\f$ and is
determined from (3.2).  We note that the ``constituent'' fields transform as
$\tilde{q}_L\Rt K\tilde{q}_L$ and $\tilde{q}_R\Rt K\tilde{q}_R$.  The vector
and
pseudovector combinations:
$$
v_{\m}=\frac{i}{2}(\x\pa_{\m} \x^{\dag}+\x^{\dag}\pa_{\m}\x)
$$
\begin{equation}
p_\m=\frac{i}{2}(\x\pa_\m\x^{\dag}-\x^{\dag}\pa_{\m}\x)
\end{equation}
are seen to transform as
$$
v_{\m}\Rt Kv_{\m}K^{\dag} +iK\pa_{\m}K^{\dag}
$$
\begin{equation}
p_{\m}\Rt Kp_{\m}K^{\dag}.
\end{equation}
Using (3.4) we can construct a covariant chiral derivative acting on
``constituent'' type fields:
$$
D_{\mu} \tilde{q}=(\pa_{\m}-iv_{\m})\tilde{q},
$$
\begin{equation}
D_{\m} \tilde{q}\Rt KD_{\m}\tilde{q}.
\end{equation}
These transformation properties enable us to simply construct chiral
invariants.

Before adding the heavy fields into this picture let us add the ``light''
vector fields.  There is a strong phenomenological motivation to do so, of
course.  But there is also a kind of aesthetic reason which is motivated by the
heavy quark symmetry.  This is simply that the heavy meson multiplet (2.12)
involves both pseudoscalars and vectors.  If we want to imagine models in which
we can try to extrapolate some quark masses up and down it is necessary to
include all relevant degrees of freedom.

It is straightforward to introduce {\it both} vector and axial vector mesons as
linear combinations of fields transforming like
\begin{equation}
A^L_{\m}\Rt U_LA^L_{\m}U^{-1}_L,~~~A^R_{\m}\Rt U_RA^R_{\m}U^{-1}_R.
\end{equation}
For reasons of economy (and because we are not including the other $\ell=1$
$\bar{q}q$ states) we would like to ``integrate out'' the axials, analogously
to the way one ``integrates out'' the scalar sigma meson in arriving at the
non-linear sigma model.  This can be done \cite{Kay} by writing
$A^L_\m$ and $A^R_\m$ in terms of the physical vector field $\r_\m$ (a $3\times
3$ matrix):
$$
A^L_{\m}=\x\r_{\m}\x^{\dag}+\frac{i}{\tilde{g}}\x\pa_{\m}\x^{\dag}
$$
\begin{equation}
A^R_{\m}=\x^{\dag}\r_{\m}\x +\frac{i}{\tilde{g}}\x^{\dag}\pa_{\m}\x,
\end{equation}
where $\tilde{g}$ is a vector meson coupling constant.  $\r_{\m}$ is seen to
transform as
\begin{equation}
\r_{\m}\rightarrow K\r_{\m}K^{\dag} + \frac{i}{\tilde{g}}K\pa_{\m}K^{\dag}
\end{equation}
so that $F_{\m\n}(\r)=\pa_{\m}\r_{\n}-\pa_{\n}\r_{\m}
-i\tilde{g}[\r_{\m},\r_{\n}]$ transforms as
\begin{equation}
F_{\m\n}(\r)\rightarrow KF_{\m\n}(\r)K^{\dag}.
\end{equation}
It is easy to construct chiral invariants using (3.6) and (3.9).  The
``minimal'' chiral Lagrangian of light pseudoscalars and vectors is then
simply \cite{Kay} \cite{Fuj}
$$
\cl_{\rm light}=-\frac{1}{4} Tr[F_{\m\n}(\r)F_{\m\n}(\r)]
$$
$$
-\frac{m^2_v}{8K} (1+K)Tr(A^L_{\m}A^L_{\m}+A^R_{\m}A^R_{\m})+\frac{m^2_v}{4K}
(1-K)Tr(A^L_{\m}UA^R_{\m}U^{\dag}),
$$
\begin{equation}
K=(m_v/F_{\p}\tilde{g})^2,
\end{equation}
where $m_v$ is the light vector meson mass.  Note that (3.10) also contains the
kinetic and interaction terms of the pseudoscalar mesons.  Chiral symmetry
breaking terms as well as terms proportional to $\e_{\m\n\a\b}$ are given
elsewhere. \cite{Kay} \cite{Jai}  The coupling constant $\tilde{g}$ is
related \cite{Jai} to the width $\G(\r\rightarrow 2\p)$; a suitable value is
\begin{equation}
\tilde{g}\simeq 3.93.
\end{equation}

Now consider the ``ordinary'' heavy meson fields $P$ and $Q_{\m}$ discussed in
section 2.  Under the chiral transformations only the light ``constituent''
degrees of freedom transform so (see (2.1)) we have
\begin{equation}
P\rightarrow PK^{\dag},~~~~~Q_{\m}\rightarrow Q_{\m}K^{\dag}.
\end{equation}
We can upgrade (2.2) and (2.9) to chiral invariants involving interactions with
light pseudoscalars and vectors merely by replacing the derivative operators
appearing there by suitable
covariant derivatives.  At this point, however, there is an interesting choice.
As can be seen from (3.5) and (3.8) both the vector combination of
pseudoscalars, $v_{\m}$ and the vector particles, $\r_{\m}$ transform in the
same way.  Let us therefore define a generalized covariant derivative,
$$
\cd_{\m}\bar{P}=[\pa_{\m}-i\a\tilde{g}\r_{\m}-i(1-\a)v_{\m}]\bar{P},
$$
\begin{equation}
\cd_{\m}P=P[\buildrel\leftarrow\over{\pa}_{\m}+i\a\tilde{g}\r_{\m}
+i(1-\a)v_{\m}].
\end{equation}
(The same definitions hold for $\bar{Q}_{\n}$ and $Q_{\n}$.)  The dimensionless
parameter $\a$ specifies the extent to which two emitted pseudoscalars in a
relative p-wave like to arise from an intermediate vector state; $\a=1$ would
correspond to ``vector meson dominance''.  Our prejudice is that $\a$ should be
close to unity.  However, $\a$ should eventually be found by comparing
calculations in this model to experiment.  Now, making the substitutions (2.3)
and (2.10) in the ``covariantized'' (2.2) and (2.9) yields for the small
oscillation fields
\begin{equation}
\cl^{(1)}_{\rm heavy}=iMV_{\m}Tr[H(\pa_{\m}-i\a\tilde{g}\r_{\m}-i(1-\a)v_{\m})
\bar{H}],
\end{equation}
in which terms $\co(1)$ in $M$ were neglected and (2.12) was used.  Note
that the ``Tr'' symbol pertains to the Dirac space while the light flavor space
summation is implicit.

Before discussing other chiral invariant interaction terms let us give the
hermiticity and CP transformation properties for the quantities involved.
Under hermiticity
$$
P\leftrightarrow \bar{P},~~~Q_{\m}\leftrightarrow (-1)^{\d_{\m 4}}\bar{Q}_{\m},
$$
$$
\r_{\m}\leftrightarrow (-1)^{\d_{\m 4}}p_{\m},~~~
p_{\m}\leftrightarrow (-1)^{\d_{\m 4}}p_{\m},~~~
$$
\begin{equation}
\pa_{\m}\leftrightarrow (-1)^{\d_{\m 4}}\pa_{\m},
\end{equation}
while the {\em usual} phase conventions would give the CP properties:
$$
P\leftrightarrow -\bar{P}^T,~~~Q_{\m}\leftrightarrow (-1)^{\d_{\m
4}}\bar{Q}^T_{\m},
$$
$$
\r_{\m}\leftrightarrow (-1)^{\d_{\m 4}}\r^T_{\m},~~~p_\m \leftrightarrow
(-1)^{\d_{\m 4}}p^T_{\m},
$$
\begin{equation}
\pa_{\m}\leftrightarrow -(-1)^{\d_{\m 4}}\pa_{\m}.
\end{equation}
If we want to implement an effective CP operation for the heavy primed fields
so that the same Lagrangian describes both the heavy quark and heavy antiquark
sectors we should demand that under CP:
$$
P'\leftrightarrow -\bar{P}'^T,~~~Q'_{\m}\leftrightarrow (-1)^{\d_{\m
4}}\bar{Q}'^T_{\m}.
$$
\begin{equation}
V_{\m}\iff (-1)^{\d_{\m 4}}V_{\m}
\end{equation}
Note especially that the behavior of $V_\m$ follows from requiring, for
example, that $V_{\m}P'\Rt -(-1)^{\d_{\m 4}}V_{\m}\bar{P}'^T$ to match, using
(2.3), the result for ordinary fields that $\pa_{\m}P\Rt (-1)^{\d_{\m4}}
\pa_{\m}\bar{P}^T$.  It was shown in ref. 2 that an interaction term
\begin{equation}
\cl^{(2)}_{\rm heavy}=iMd Tr(H\g_{\m}\g_5 p_{\m}\bar{H})
\end{equation}
plays an important role.  $d$ is a real dimensionless constant.  Let us expand
this, using (2.12), to find
$$
\cl^{(2)}_{\rm heavy}=2Md[\h P'p_{\m}\bar{Q}'_{\m}+\h^*Q'_{\m}p_{\m}\bar{P}
$$
\begin{equation}
+\e_{\b\a\r\m}V_{\r}Q'_{\a}p_{\m}\bar{Q}'_{\b}].
\end{equation}
As it stands (3.18) is hermitean.  But, using the mnemonics in (3.17) and
(3.16) we see that the first term goes to $-\h Q'_{\m} p_{\m}\bar{P}'$ under
CP.
In order to agree with the second term we require $\h=-\h^*$, which is
satisfied by choosing $\h=i$.  It may be verified that the third term is also
CP invariant.  We note that (3.19) is descended from the ordinary field
Lagrangian
$$
\cl^{(2)}_{\rm heavy}(P,Q)=2iMd [Pp_{\m}\bar{Q}_{\m}-Q_{\m}p_{\m}\bar{P}
$$
\begin{equation}
-\frac{1}{2M} \e_{\b\a\r\m}
(\cd_{\r}Q_{\a}p_{\m}\bar{Q}_{\b}-Q_{\a}p_{\m}{\cd}_{\r}\bar{Q}_{\b})].
\end{equation}
The heavy quark symmetry has related the coefficients of the two different
pieces in (3.20).

Another important chiral invariant interaction may be written in the heavy
symmetry limit
$$
\cl^{(3)}_{\rm heavy}=\frac{icM}{m_v} Tr(H\g_{\m}\g_{\n}F_{\m\n}(\r)\bar{H})
$$
\begin{equation}
=\frac{-2cM}{m_v}[2iQ'_{\m}F_{\m\n}(\r)\bar{Q}'_{\n}+\e_{\m\n\a\b}V_{\b}
(P'F_{\m\n}(\r)\bar{Q}'_{\a}-Q'_{\a}F_{\m\n}(\r)\bar{P}')],
\end{equation}
where $c$ is a dimensionless constant and the light vector mass $m_{\n}$
appears just for dimensional reasons.  Equation (3.21) is the limit of the
ordinary field Lagrangian,
$$
\cl^{(3)}_{\rm heavy}(P,Q)=\frac{2icM}{m_v}\left[-2Q_{\m}F_{\m\n}
(\r)\bar{Q}_{\n}\right.
$$
\begin{equation}
\left.+ \frac{1}{M}
\e_{\m\n\a\b}(\cd_{\b}PF_{\m\n}(\r)\bar{Q}_{\a}+Q_{\a}F_{\m\n}
(\r)\cd_{\b}\bar{P})\right] .
\end{equation}
Again, CP invariance may be verified and heavy quark symmetry is seen to have
related the coefficients of the two pieces.  We can also construct a term
similar to (3.21) or (3.22) in which $F_{\m\n}(\r)$ is replaced by
$F_{\m\n}(v)=\pa_{\m} v_{\n}-\pa_{\n}v_{\m}-i[v_{\m},v_{\n}]$:
\begin{equation}
\cl^{(3)'}_{\rm heavy} =\frac{ic'M}{F_\p}
Tr(H\g_{\m}\g_{\n}F_{\m\n}(v)\bar{H}).
\end{equation}
In the spirit of (light) vector meson dominance we would expect this term to be
less important.  However it would play a role in a model where vectors are
neglected.

To sum up, the leading terms of the chiral invariant heavy meson Lagrangian
written in terms of the doublet $H$ field are
\begin{equation}
\cl_{\rm heavy}~=~(3.14)~+~(3.18)~+~(3.21).
\end{equation}
These involve the new coupling constants $c$ and $d$ as well as the parameter
$\a$ which would be unity in the vector meson dominance approximation.  We will
not explicitly write the chiral symmetry breaking terms here.  Strictly
speaking, (3.24) is defined only for mesons with a heavy {\em quark}.  A
continuation of (3.24) to ``ordinary'' heavy fields (containing also meson
states with heavy {\em antiquarks}) is provided by
\begin{equation}
\cl(P,Q)=[(2.2)~+~(2.9)]{\rm with} \pa_{\m}\Rt \cd_{\m}~+~(3.20)~+~(3.22).
\end{equation}
The mutual consistency of (3.24) and (3.25) led to the determination of the
phase $\h$ in the definition of $H$, (2.12). These Lagrangians should be used
for large $M$ and small momenta (more precisely, small $p\cdot V$) of the light
pseudoscalars and vectors.
\sxn{Weak currents}
One of the main applications \cite{Wis} of the heavy quark approach is to the
semi-leptonic decays of heavy mesons.  These are governed by the effective weak
interaction
$$
\cl_W = \frac{G_F}{\sqrt 2} J^{(+)}_{\m}(x)J^{(-)}_{\m}(x),
$$
$$
J^{(-)}_{\m}=i\sum_{\ell,k}\bar{u}_{\ell}V_{\ell k}\g_{\m}(1+\g_5)d_k
+i\bar{\n}_e \g_{\m}(1+\g_5)e +\cdots,
$$
\begin{equation}
J^{(+)}_{\m}=(-1)^{\d_{\m4}} J^{(-){\dag}}_{\m},
\end{equation}
wherein $G_F$ is the Fermi constant, $V_{\ell k}$ is the quark mixing
(Kobayashi Maskawa) matrix, $u_l$ is the $\ell^{th}$ charge 2/3 quark, $d_k$ is
the $k^{th}$ charge -1/3 quark and other leptonic terms (as well as possible
lepton mixings) have not been included.  Note that if the matrix elements of
$V_{\ell k}$ were all real, $J_{\m}^{(-)}$ would go to $J_{\m}^{(+)}$ under CP
and $\cl_W$ would be CP invariant.  The hadronic currents of immediate interest
are
$$
J^{(-)}_{\m}=iV_{cs}\bar{c}\g_{\m}(1+\g_5)s+iV_{ub}\bar{u}\g_{\m}
(1+\g_5)b+\cdots,
$$
\begin{equation}
J^{(+)}_{\m}=iV^*_{cs}\bar{s}\g_{\m}(1+\g_5)c+iV_{ub}^*\bar{b}\g_{\m}(1+\g_5)
u+\cdots.
\end{equation}
In this paper we will confine our attention to hadronic transitions (current
matrix elements) of the form heavy meson $\Rt$ light mesons.  We need the
realization of the operator $i\bar{q}_a\g_{\m}(1+\g_5)q_{\rm heavy}$, where
$q_a$ is a light quark, in terms of meson fields.  The normalizations are
provided by the matrix elements for the pseudoscalar $\Rt$ vacuum transitions:
$$
i\bar{q}_a\g_{\m}(1+\g_5)q_{\rm heavy}=F\pa_{\m}P_a+\cdots ,
$$
\begin{equation}
i\bar{q}_{\rm heavy}\g_{\m}(1+\g_5)q_a=F\pa_{\m}\bar{P}_a+\cdots ,
\end{equation}
where $F$ is the decay constant for each particular heavy meson.  (In this
normalization convention $F_\p\simeq$ 132 MeV).  Heavy quark symmetry gives the
normalization for the heavy vector $\Rt$ vacuum transition in terms of $F$.
The leading order chiral covariant generalization was already presented in ref.
2.  In our notation it reads
$$
i\bar{q}_a\g_{\m}(1+\g_5)q_{\rm heavy}=\frac{-iFM}{2}
Tr[\g_{\m}(1+\g_5)H_b]\x^{\dag}_{ba}+\cdots ,
$$
\begin{equation}
i\bar{q}_{\rm heavy}\g_{\m}(1+\g_5)q_a
=\frac{-iFM}{2}\x_{ab}Tr[\g_{\m}(1+\g_5)\bar{H}_b]+\cdots ,
\end{equation}
wherein all fields are being evaluated at $x_{\m}=0$ (to eliminate a phase
which would arise when (4.4) is derived using (2.3), (2.10) and (2.12)).  The
factor of $\xi$ in the second equation, for example, is required for chiral
covariance: The quark current on the LHS transforms with a factor $U_L$.  On
the other hand $\bar{H}\Rt K\bar{H}$.  Using (3.2) it is seen that
$\xi\bar{H}\Rt U_L(\xi\bar{H})$.  Eq. (4.4) gives the currents in the heavy
quark limit.  It is descended from (using (2.3), (2.10) and neglect of
subleading terms in $M$) the ``ordinary field'' currents
$$
i\bar{q}_a\g_{\m}(1+\g_5)q_{\rm heavy}
=F(\cd_{\m}P_b+MQ_{b\m})(\xi^{\dag})_{ba}+\cdots ,
$$
\begin{equation}
i\bar{q}_{\rm heavy}\g_{\m}(1+\g_5)q_a
=F\xi_{ab}(\cd_{\m}\bar{P}_b+M\bar{Q}_{b\m})+\cdots .
\end{equation}
Note that the covariant derivatives include pieces proportional to $\r_{\m}$
which are formally suppressed by $\frac{1}{M}$.

Without necessarily endorsing the notions that the $c$ quark is truly heavy
and the $s$ quark is truly light we give a specific example of a term in (4.2)
and (4.5):
\begin{equation}
J^{(+)}_{\m}=V^*_{cs}F_D[\pa_{\m}D_b+i\a\tilde{g}D_c\r_{{\m}cb} +
i(1-\a)v_{{\m}cb}+M_DD^*_{{\m}b}](\xi)^{\dag}_{\b3}+\cdots ,
\end{equation}
where $D_b=(D^0,D^+,D^+_s)$ and $D^*_{{\m}b}$ denotes the vector triplet field.
We will present results for the $D\Rt \bar{K}$ and $D\Rt \bar{K}^*$ transitions
based on (4.6).  Essentially identical formulas will hold for the $B\Rt \p$ and
$B\Rt \r$ transitions, etc.
\sxn{Applications}
For orientation let us first consider the hadronic matrix element for the decay
$D^0\Rt K^-e^+\n_{e}$, even though it is practically the same as that for
$\bar{B}^0\Rt \p^+e^-\n_e$, already discussed. \cite{Wis} \cite{Bar}  The
invariant matrix element is parametrized by
\begin{equation}
\sqrt{4p_0p'_0V^2}<K^-(p')|J^{+}_{\m}|D^0(p)>
=V^*_{cs}[f_+(q^2)(p+p')_{\m}+f_-(q^2)(p-p')_{\m}],
\end{equation}
where $q=p-p'$.  There is first of all, a direct transition which is read off
from (4.6) and (3.3) to be $V^*_{cs}\frac{F_D}{F_\p}p_{\m}$.  In addition,
there is a contribution from the $D^{*+}_s$ pole diagram.  This has three
factors: $V^*_{cs}MF_D$ from (4.6), $2iMdp'_{\n}/F_\p$ from (3.20) and the
usual $D^{*+}_s$ propagator.  Putting everything together gives:
$$
f_++f_-=\frac{F_D}{F_\p}\left[1+\frac{2dM^2p'\cdot
q}{M^{*2}_s}~\frac{1}{q^2+M^{*2}_s}\right],
$$
\begin{equation}
f_+-f_-=2dM^2\frac{F_D}{F_{\p}}~\frac{1-p'\cdot q/M^{*2}_s}{q^2+M^{*2}_s}.
\end{equation}
These formulas are expected to be valid only for soft kaons, $p'\cdot V$ small
where $V_{\m}=p_{\m}/M$.  They may be simplified by writing
$q_{\m}=MV_{\m}-p'_{\m}$ and formally neglecting terms $\co(p'^2)$ to yield
\cite{Wis}
$$
f_++f_-=\frac{F_D}{F_\p}\left[1+\frac{dV\cdot p'}{\D-p'\cdot V}\right],
$$
\begin{equation}
f_+-f_-=\frac{F_D}{F_\p}\frac{dM}{\D-p'\cdot V},
\end{equation}
where $\D=M^*_s-M$.  We may check the large $M$ scaling laws which say
\cite{Wis} that (5.1) should be of order $M^{1/2}$.  $f_++f_-$ is the
coefficient of $MV_{\m}$ so it should go as $M^{-1/2}$ which it does because
$F_D \sim M^{-1/2}$.  On the other hand $f_+-f_-$ is the coefficient of
$p'_{\m}$ so it should go as $M^{1/2}$.  That it does so is most evident from
(5.3).  Note that in the extreme $M\Rt \infty$ limit we thus expect $f_++f_-$
to
vanish.  In this limit
\begin{equation}
f_+(q^2)\sim \frac{dM^2F_D/F_\p}{q^2+M^{*2}_s}
\end{equation}

We should stress that (5.4) is theoretically justified only near $q^2=-M^2$;
there is no reason for it to hold near $q^2=0$ in the present approach.
Nevertheless, Anjos et.al. \cite{Anj} find that such a global form fits their
experiment with $f_+(0)=0.79\pm 0.05\pm 0.06$.  This would imply
\begin{equation}
d\frac{F_D}{F_\p} \approx 1,
\end{equation}
which may perhaps be safely interpreted as giving the rough order of magnitude
of $d$ (Wise \cite{Wis} finds $|d|<1.7$).

Now let us turn to the matrix element
\begin{equation}
\sqrt{4p_0p'_0V^2}<\bar{K}^{*0}(p',\e)|J^{(+)}_{\m}/V^*_{cs}|D^+(p)>,
\end{equation}
where $\e$ is the $\bar{K}^{*0}$ polarization vector, which is relevant for
$D^+\Rt \bar{K}^{*0}e^+\n_e$.  One would of course expect the transition $B\Rt
\r$ to be a case which is better approximated in our approach.  The predicted
formulas would be the same but we have chosen the case in (5.6) because
experimental data exist for it.  According to the large $M$ scaling rules,
(5.6) behaves as $M^{1/2}$, just like (5.1).  [Actually, with our state
normalization convention, this can be read off from the external heavy meson
factor $\sqrt{p_0}$ on the LHS].  There are several contributions to (5.6) from
the leading order current operator (4.6) together with the heavy Lagrangian
(3.25).  First there is a direct contribution from (4.6):
\begin{equation}
i\a\tilde{g}F_D\bar{\e}_{\m}
\end{equation}
where $\bar{\e}_{\m}=(-1)^{\d_{\m4}}\e^*_{\m}$.  However this formally goes as
$M^{-1/2}$ and must be interpreted as vanishing in the strict heavy quark
limit.  There is also a $D^+_s$ pole contribution which is found from (4.6) and
the covariantized heavy kinetic term to be:
\begin{equation}
-2i\a\tilde{g}F_D~\frac{p\cdot \bar{\e}q_{\m}}{q^2+M^2_s},
\end{equation}
where $q_{\m}=p_{\m}-p'_{\m}$.  Because of the overall $q_{\m}$ this term will
not contribute to physical processes to the extent that the lepton current is
conserved (where $m_e$ is negligible).  It may, however, be someday measured in
$D^+\Rt \bar{K}^{*0}\m^+\n_{\m}$.  Its heavy quark limit is obtained by setting
$p_{\m}=MV_{\m}$, $q_{\m}=MV_{\m}-p'_{\m}$ and considering $p'_{\m}$ small; the
result
\begin{equation}
-i\a\tilde{g}F_DM\frac{V\cdot \bar{\e}V_{\m}}{(M_s-M)-p'\cdot V}
\end{equation}
immediately shows that (5.8) properly scales as $M^{1/2}$.  The remaining
leading contribution to (5.6) is of vector rather than axial-vector type; using
(4.6) and (3.22) we find that the $D^{*+}_s$ pole diagram gives
\begin{equation}
4icF_D(M/m_v)\e_{\s\n\m\b}\frac{p_{\b}p'_\s\bar{\e}_{\n}}{q^2+M^{*2}_s},
\end{equation}
where $m_v$ is the light vector mass and $c$ is the new coupling constant
introduced in (3.22).  Its heavy quark limit,
\begin{equation}
2icF_D(M/m_v)\e_{\s\n\m\b}\frac{V_{\b}p'_{\s}\bar{\e}_\n}{(M^*_s-M)-p'\cdot V}
\end{equation}
is seen to scale as $M^{1/2}$ (with $c$ scaling as $M^0$).

To sum up, our result for the $D\Rt \bar{K}^*$ transition matrix element is
given by adding (5.7), (5.8) and (5.10).  It should be used only for $q^2$ near
$-M^2$.  In particular, there is no justification for using it near $q^2=0$.

The experimental situation, which has recently been reviewed by Pham
\cite{Pha}, is a little complicated since three different form factors (one
other proportional to $p_{\m}$ in addition to those in (5.7) and (5.10)) must
be inferred from the data.  He shows that the coefficient of $\bar{\e}_{\m}$ is
most crucial in establishing the overall rate; with a phenomenological $q^2$
damping the coefficient of $\bar{\e}_{\m}$ at $q^2=0$ is found to be roughly
1.4 GeV in magnitude.  This might be contrasted with our (5.7) which (for
$\a=1$) has a magnitude of about 0.6 GeV.  Note that (5.7) is only expected to
be valid near $q^2=-M^2$.  Typically the form factors fall with increasing
$-q^2$ so our result does not seem unreasonable.  Since (5.7) vanishes in the
$M\Rt \infty$ limit, this also provides a caution about using the infinite $M$
limit for the $D$ meson.  Continuing, Pham finds that the coefficient of
$p_{\m}$ in (5.6) is perhaps consistent with zero.  This feature would also
agree with our result.  Finally, the coupling constant $c$ could be eventually
determined by comparison of the experimental vector-type form factor in the
$q^2\approx -M^2$ region with (5.10).

All in all, the use of the chiral symmetric expression (5.7) + (5.8) + (5.10)
for small light vector momenta does not seem to be unreasonable at our present
stage of experimental understanding of the process $D^+\Rt
\bar{K}^{*0}e^+\n_e$.  For $D$ decays, at least, it seems better to use the
expression (5.7) which follows from the ``ordinary'' field Lagrangian rather
than its strict $M\Rt \infty$ limit (of zero).  In the future it would be
interesting to include terms non-leading in $M$ and also terms containing more
derivatives of the light fields.  An example of the latter which has a piece
giving a direct (non-pole) heavy pseudoscalar $\Rt$ light vector transition is
(cf (4.5)):
\begin{equation}
i\bar{q}_a\g_{\m}(1+\g_5)q_{\rm heavy}=\cdots
+i\b\cd_{\n}P_c[F_{\m\n}(\r)]_{cb}[\x^{\dag}]_{ba}+\cdots,
\end{equation}
where $\b$ is a real constant.  This yields the extra contribution to the
matrix element (5.6):
\begin{equation}
i\b[p\cdot \bar{\e}(p_\m-q_\m)+p\cdot p'\bar{\e}_\m].
\end{equation}
This is seen to contain a piece proportional to $p_\m$ which was absent from
our leading order result above.
\sxn{Summary}
We have given the general framework for adding ``light'' vector particles to
the heavy hadron effective chiral Lagrangian.  The leading order in $M$ strong
Lagrangian was presented in section 3 with special attention to a convenient
phase convention for including the anti-quark sector.  There are many possible
applications, the most immediate being to the semi-leptonic decays of heavy
mesons.  We discussed the process $D^+\Rt \bar{K}^{*0}e^+\n_e$ (chosen because
data exists) and found that the leading order result was reasonable at our
present stage of experimental knowledge.  In the future it would be instructive
to compare the results with those computed for $D^+\Rt K^-\p^+e^+v_e$.  Many
other processes with (extra) ``soft'' light vectors and/or pseudoscalars can be
considered.  Decays like $B\Rt D{\r}e\bar{\n}_e$ are very natural to look at
next.
Progress along these and similar lines will be reported elsewhere.

\noindent{\bf Acknowledgements}

We would like to thank B.S. Balakrishna, K. Gupta, S. Rajeev and Z. Yang for
helpful discussions.  This work was supported in part by the U.S. Department of
Energy under contract number DE-FG-02-85ER40231.

\br
\rf{Wis} A review with complete references is furnished by M. Wise, Lectures
at the Lake Louis Winter Institute, Feb. 17-23, 1991.
\rf{Wi2} M. Wise, Phys. Rev. {\bf D45}, R2188 (1992).
\rf{Yan} T.M. Yan, H.Y. Cheng, C.Y. Cheung, G.L. Lin, Y.C. Lin and H.L. Yu,
Phys. Rev. {\bf D46}, 1148 (1992).
\rf{Bar} G. Burdman and J.F. Donoghue, Univ. of Massachusetts report
UMHEP-365.
\rf{Cho} P. Cho, Harvard report HUTP-92/A014.
\rf{Jen} E. Jenkins, A.V. Manohar and M. Wise, Cal Tech report
CALT-68-1783.
\rf{Par} See note v following the meson summary table in Review of
Particle Properties, Physical Review {\bf D45}, Part 2 (June 1992).
\rf{Geo} H. Georgi, Phys. Lett {\bf B240}, 447 (1990).
\rf{Bjo} J.D. Bjorken, SLAC report SLAC-PUB 5278.
\rf{Cro} J. Cronin, Phys. Rev. {\bf 161}, 1483 (1967).
\rf{Cal} C. Callan, S. Coleman, J. Wess and B. Zumino, Phys. Rev. {\bf 177},
2249 (1969).
\rf{Kay} \"{O}. Kaymakcalan and J. Schechter, Phys. Rev. {\bf D31}, 1109
(1985).
\rf{Fuj} A different approach given by T. Fujiwara, T. Kugo, H. Terao, S.
Uehara and K. Yamawaki, Prog. Theor. Phys. {\bf 73}, 926 (1985) leads to the
identical Lagrangian.
\rf{Jai} A number of detailed investigations show that the light
pseudoscalar-vector chiral Lagrangian can explain many features of the low
energy dynamics and gives a much improved description of the soliton sector
compared to the Lagrangian with pseudoscalars only.  See P. Jain, R. Johnson,
Ulf-G. Meissner, N.W. Park and J. Schechter, Phys. Rev. {\bf D37}, 3252 (1988);
Ulf-G. Meissner, N. Kaiser, H. Weigel and J. Schechter, Phys. Rev. {\bf D39},
1956 (1989); P. Jain, R. Johnson, N.W. Park, J. Schechter and H. Weigel, Phys.
Rev. {\bf D40}, 885 (1989); B. Schwesinger, H. Weigel, G. Holzwarth and A.
Hayashi, Phys. Rep. {\bf 173}, 173 (1989);
J. Schechter, V. Soni, A. Subbaraman and H. Weigel,
Mod. Phys. Lett. {\bf A7}, 1 (1992); N.W. Park and H. Weigel, Nucl. Phys.
{\bf A541}, 453 (1992).  An updating of the symmetry breaking in this model
will
appear soon: J. Schechter, A. Subbaraman and H. Weigel, forthcoming
Syracuse-Turbingon report.
\rf{Anj} J.C. Anjos et.al., Phys. Rev. Lett. {\bf 62}, 1587 (1989).
\rf{Pha} X.-Y. Pham, Laboratoire de Physique Th\'{e}orique et Haute Energies,
Paris report LP THE 92-12.
\end{thebibliography}

\end{document}